\documentclass[aps,prb,twocolumn,superscriptaddress,showpacs]{revtex4}
\usepackage{latexsym}
\usepackage{graphicx}
\usepackage{graphicx}% Include figure files
\usepackage{dcolumn} % Align table columns on decimal point
\usepackage{bm}
\usepackage{epsfig}
\begin{document} 

\title{Fermiology and transport properties of the half-metallic itinerant ferromagnet CoS$_2$: influence of spin orbit coupling}

\author{A. Pi\~neiro}
\email{alberto.pineiro@usc.es}
\affiliation{Departamento de F\'{i}sica Aplicada,
  Universidad de Santiago de Compostela, E-15782 Campus Sur s/n,
  Santiago de Compostela, Spain}
\affiliation{Instituto de Investigaciones Tecnol\'{o}gicas,
  Universidad de Santiago de Compostela, E-15782 Campus Sur s/n,
  Santiago de Compostela, Spain}
\author{A.S. Botana}
\affiliation{Departamento de F\'{i}sica Aplicada,
  Universidad de Santiago de Compostela, E-15782 Campus Sur s/n,
  Santiago de Compostela, Spain}
\affiliation{Instituto de Investigaciones Tecnol\'{o}gicas,
  Universidad de Santiago de Compostela, E-15782 Campus Sur s/n,
  Santiago de Compostela, Spain}  
\author{V. Pardo}
\affiliation{Departamento de F\'{i}sica Aplicada,
  Universidad de Santiago de Compostela, E-15782 Campus Sur s/n,
  Santiago de Compostela, Spain}
\affiliation{Instituto de Investigaciones Tecnol\'{o}gicas,
  Universidad de Santiago de Compostela, E-15782 Campus Sur s/n,
  Santiago de Compostela, Spain}
\author{D. Baldomir}
\affiliation{Departamento de F\'{i}sica Aplicada,
  Universidad de Santiago de Compostela, E-15782 Campus Sur s/n,
  Santiago de Compostela, Spain}
\affiliation{Instituto de Investigaciones Tecnol\'{o}gicas,
  Universidad de Santiago de Compostela, E-15782 Campus Sur s/n,
  Santiago de Compostela, Spain}
\date{\today}

\begin{abstract}
Electronic structure calculations were performed on the compound CoS$_2$, an itinerant ferromagnet whose magnetic properties can be understood in terms of spin fluctuation theory. We have identified nesting features in the Fermi surface of the compound, active for long wavelength spin fluctuations. The electronic structure of the material is close to a half-metal. We show the importance of introducing spin-orbit coupling (SOC) in the calculations, that partially destroys the half-metallicity of the material. By means of transport properties calculations, we have quantified the influence of SOC in the conductivity at room temperature, with an important decrease comparing to the GGA alone conductivity. SOC also helps to understand the negative 0 of the material, whose conductivity varies by a few percent with the introduction of small perturbations in the states around the Fermi level.
\end{abstract}
\maketitle

\section{Introduction}

Half metals\cite{katsnelson, schwarz_half} are materials where metallic conduction occurs only through one of the spin channels, leading to a perfectly polarized electron current. One of the spin channels presents a gap at the Fermi Energy (E$_F$) but the other one has bands crossing it. This makes them an interesting family of materials, in particular for applications in the field of spintronics,\cite{spintronic} where spin injection is required for controlling the charge and spin currents separately. Moving a strongly spin-polarized current is highly sought for, and half-metals are a good candidate for this.

Several electronic structure calculations based on density-functional theory (DFT) have been done in the last few years, describing CoS$_{2}$ as a half-metallic ferromagnet.\cite{zhao, mazin, kwon, tatsuya, jiu_jin, ning_wu} However, the electronic and magnetic properties of this material are not fully understood. Point-contact Andreev reflection measurements\cite{de_jong} showed a relatively low-spin polarization of 56\% at $4.2$K.\cite{wang} This picture was confirmed in the studies by Wang \textit{et al.}\cite{wang_2} in the $Co_{1-x}Fe_{x}S_{2}$ series, where they show that the spin polarizations can be 0 tuned in the range $-56\%<P<+85\%$. Brown \textit{et al.}\cite{brown} found from polarized neutron-diffraction measurements that half metallicity does not occur in the ferromagnetic phase. A sharp photoemission peak at the Fermi energy in the ferromagnetic phase originates in the bottom of $e_{g\downarrow}$ subband due to the exchange 3, indicating that the $e_{g\downarrow}$ band is partially filled in the ferromagnetic phase,\cite{takahashi} destroying the half metallicity.

Recently it was found that the resistivity, specific heat and magnetic susceptibility of CoS$_{2}$ is dominated by exchange-enhanced spin-density fluctuations.\cite{martaprb} This previous work also show that the GGA method\cite{gga} is enough to describe the electronic and magnetic 0 of CoS$_{2}$ due to the non localized nature of the compound. There is no need to introduce strong correlation effects to describe accurately its electronic structure properties but not enough to describe the enhancement of, e.g. the value $\gamma$ obtained from specific heat , which is enhanced due to spin fluctuations unaccounted for in DFT calculations.

The important role of spin-orbit coupling in a realistic description of half-metallic materials has been discussed in literature about half-metals.\cite{katsnelson,mavropoulos} Usually the coupling between spin-up and spin-down states caused by a sizable spin-orbit coupling introduces non-negligible minority states at the Fermi level, destroying the half-metallicity. 
%Mavropoulos \textit{et al.}\cite{mavropoulos} study how the effects of spin-orbit coupling change the degree of polarization of the density of states, showing the importance of the position of the band edges with respect to the Fermi level, together with the size of the interaction, marked by the atomic number of the atoms in the compound. 
In general, this will depend on the strength of the interaction (that increases with atomic size) and also on the size of the gap and the position of the band edges in the insulating spin channel.
The importance of minimizing the effects of spin-orbit coupling on destroying half-metallicity when designing new half-metals has been discussed in the past,\cite{pardo_hmafm} where utilizing cations with completely filled shells was suggested in order to reduce the spin-orbit effects by minimizing the orbital angular momenta and, in the case of studying possible compensated half-metals, increasing the likelihood of the materials having no net magnetization.

However, most works based their studies solely on the changes in the density of states at the Fermi level, but the degree of half-metallicity will be influenced by the changes in the conduction properties that derive from the reordering of states around the Fermi level introduced by spin-orbit coupling. Moreover, the conductivity of the minority spin channel will have an activated component that needs to be quantified beyond the value of the density of states at E$_F$. In this paper,  we will make an analysis of the electronic and magnetic properties of the ground state of CoS$_{2}$ and introduce spin orbit coupling (SOC) to determine its importance in the half-metallicity of the material. We will also focus on explaining, from a band structure point of view, what features help us understand its behavior as an itinerant ferromagnet, governed by spin fluctuations. For analyzing the conduction properties of the material, its spin-dependence and the importance of spin-orbit coupling, we have made several transport properties calculations and have compared them with experimental measurements.

\section{Computational details}

Electronic structure calculations were  performed within density functional theory\cite{dft} using {\sc wien2k} software,\cite{wien, wien2k} which utilizes an augmented plane wave plus local orbitals (APW+lo)\cite{sjo} method to solve the Kohn-Sham equations. This method uses an all-electron, full-potential scheme that makes no shape approximation to the potential or the electron density. The exchange-correlation potential utilized was the Perdew, Burke and Ernzerhof (PBE) version of the general gradient 0 (GGA).\cite{gga} The geometry optimization was carried out minimizing the forces in the atoms and the total energy of the system. The parameters of our calculations depend on the type of calculation but for any of them we converged with respect to the \textit{k}-mesh and to $R_{mt}K_{max}$. Values used of the \textit{k}-mesh are 7 $\times$ 7 $\times$ 7 sampling of the full Brillouin zone for electronic structure calculations and geometry optimization, 40 $\times$ 40 $\times$ 40 for the band structure calculations and the very fine details of the influence of spin orbit coupling in the half metallicity of the compound. $R_{mt}K_{max}=6.0$ is chosen for all the calculations.  Local orbitals were added for a bigger flexibility in dealing with the semicore states. Muffin tin radii chosen were the following: 2.31 a.u. for Co and 2.04 a.u. for S. 
All the calculations were converged with respect to all the parameters used, to the precision necessary to support our calculations (converged forces, total energy differences, etc.). For the calculations of transport properties we utilize the BoltzTraP code,\cite{boltztrap} that uses the energy bands obtained from the {\sc wien2}k software. We used a \textit{k}-mesh in a 43 $\times$ 43 $\times$ 43 sampling of the full Brillouin zone, where convergence was achieved. The calculation of transport properties requires a fine mesh to carry out the integrations in the Brillouin zone. Spin orbit coupling (SOC) was introduced in a second-variational procedure.\cite{singh}

\section{Structure}

\begin{figure}
\includegraphics[width=\columnwidth,draft=false]{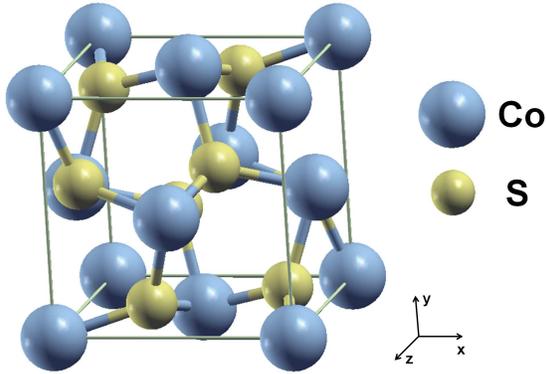}
\caption{(Color online) Cubic pyrite structure of the CoS$_2$ compound. Large blue spheres represent Co atoms and small yellow spheres represent S atoms.}\label{struct}
\end{figure}

CoS$_2$ crystallizes in a cubic pyrite structure,\cite{zhao} as shown schematically in Fig. \ref{struct}. Co and S atoms are located respectively at the Wyckoff positions 4a (0,0,0) and  8c (u,u,u) of the space group Pa$\bar{3}$, no. 205. Co atoms are octahedrally coordinated by 6 S neighbors. 

From X-ray diffraction experiments on large single crystals (3 $\times$ 3 $\times$ 3 mm$^{3}$),\cite{martaprb} we obtained the lattice parameter of the material to be $a$= 5.518 \AA. Also, electronic structure calculations were performed to optimize \textit{ab initio} the structural free parameter 0 the S position. This was calculated in order to minimize interatomic forces and the total energy of the system. The value obtained was u= 0.387 (see Table \ref{table1}), in good agreement with previous works.\cite{wyck} Calculations to obtain the lattice parameter \textit{ab initio} were performed yielding a lattice parameter \textit{a} = 5.55 \AA, which is within the typical accuracy of density functional theory calculations (2\% in volume).
 
\begin{table}[h!]
\caption{Crystallographic positions at $T=0$ that result from our structure optimization.}\label{wyc_positions}
\begin{center}
\begin{tabular}{c c c c c}
\hline
\hline
 & & & Crystallographic& \\
Atom & & & position  & Coordinates \\
\hline
Co & & & 4a & (0,0,0) \\
S & & & 8c & (0.387,0.387,0.387) \\
\hline
\end{tabular}
\end{center} 
\label{table1}
\end{table}

\section{Results}

\subsection{Electronic structure and magnetism}

\begin{figure}
\includegraphics[width=\columnwidth,draft=false]{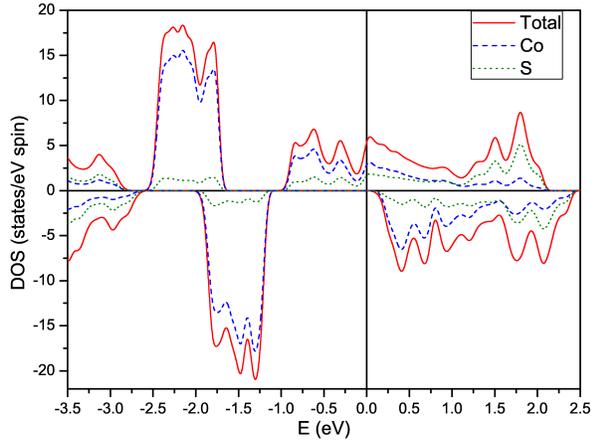}
\caption{(Color online) Spin-up (positive values) and spin-down (negative values) total density of states (DOS) plots for the CoS$_2$ in the ground state ferromagnetic configuration. Partial density of states have been plotted also (dash blue lines for Co total DOS and dot green lines for S total DOS). Fermi energy is at zero. The system is half-metallic.}\label{dos}
\end{figure}

The electronic structure of CoS$_2$ is characterized by a large covalency due to the extended S p bands overlapping largely the Co d bands close to the Fermi level. Also, the itineracy of the compound helps in making difficult to establish an ionic picture for the material. Some earlier reports suggest an effective Co$^{2+}$:d$^7$ state,\cite{tatsuya} distinct from the purely ionic picture.

Our results show that the orbital occupations are close to the Co$^{2+}$:d$^7$ state. As expected, a large covalency can be seen in the density of states (DOS) plots in Fig. \ref{dos}. Big spectral weight from S p states appears near $E_{F}$ and also significant weight can be seen at the top of the conduction bands, showing the S p bands are not fully occupied. In this DOS plot, calculated within the GGA approach, we see CoS$_2$ presents a half-metallic character, as noted in several previous works.\cite{yamamoto,tatsuya} It shows the fully occupied t$_{2g}$ bands of about 0.7 eV bandwidth at about -1 to -2 eV, that are spin-split by about 0.6 eV, and the broad e$_g^{\uparrow}$ bands crossing the Fermi level in the majority channel, being 2 eV wide. The e$_g^{\downarrow}$ bands are unoccupied, with its edge lying very close to the Fermi level. This proximity will make their occupancy non-zero under perturbations such as spin-orbit coupling, temperature or the application of an external magnetic field. A small amount of electron doping would also be able to populate them.

\begin{figure}
\includegraphics[width=\columnwidth,draft=false]{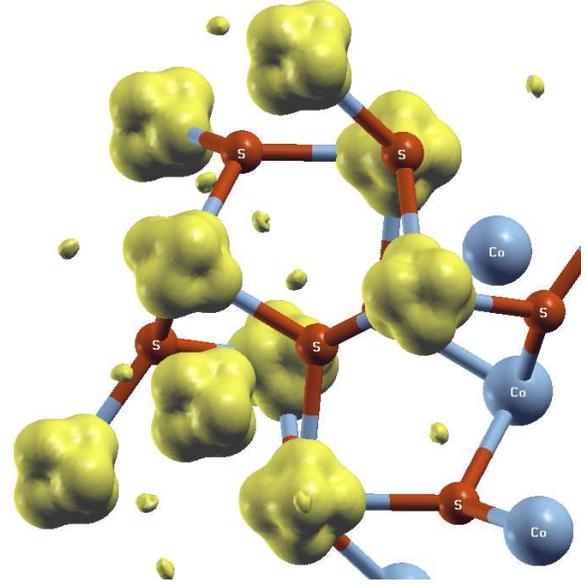}
\caption{(Color online) Charge difference of the majority and minority states of CoS$_2$ compound. Large blue spheres represent Co atoms and orange small spheres represent S atoms. The charge difference plotted correspond to an isosurface at 0.05 e/\AA$^3$ produced using XCrySDen.\cite{xcrysden} The shape of the spin density isosurface represents the orbital that carries the spin moment, being a mixture of e$_g$ states.}\label{chargue}
\end{figure}

The value of the total moment obtained by ab initio calculations is $1.00$ $\mu_B$/Co in the ground state ferromagnetic solution, in agreement with the value of the saturation magnetization found experimentally, of about 0.9 $\mu_B$/Co.\cite{martaprb} The moment comes from the partially filled majority e$_g$ band. To illustrate this description of the electronic structure, we have produced a differential electron spin density plot (Fig. \ref{chargue}). This shows the difference in electronic density between the majority and minority states of the compound (yellow), in order to see the spatial distribution of the only unpaired electron (charge difference plot will show basically the density that corresponds to the $e_{g}^{\uparrow}$ electron, because the t$_{2g}$ bands are completely full). Charge accumulation can be seen along the Co-S direction (like d$_{3z^{2}-r^{2}}$ or d$_{x^{2}-y^{2}}$ in an octahedral environment). The large bandwidth of the e$_g$ bands (about 2 eV), much higher than the crystal field splitting between the e$_g$ sub-levels, leads to a similar occupation of both e$_g$ orbitals, as we see in Fig. \ref{chargue}.

\begin{figure*}
\includegraphics[width=\textwidth,draft=false]{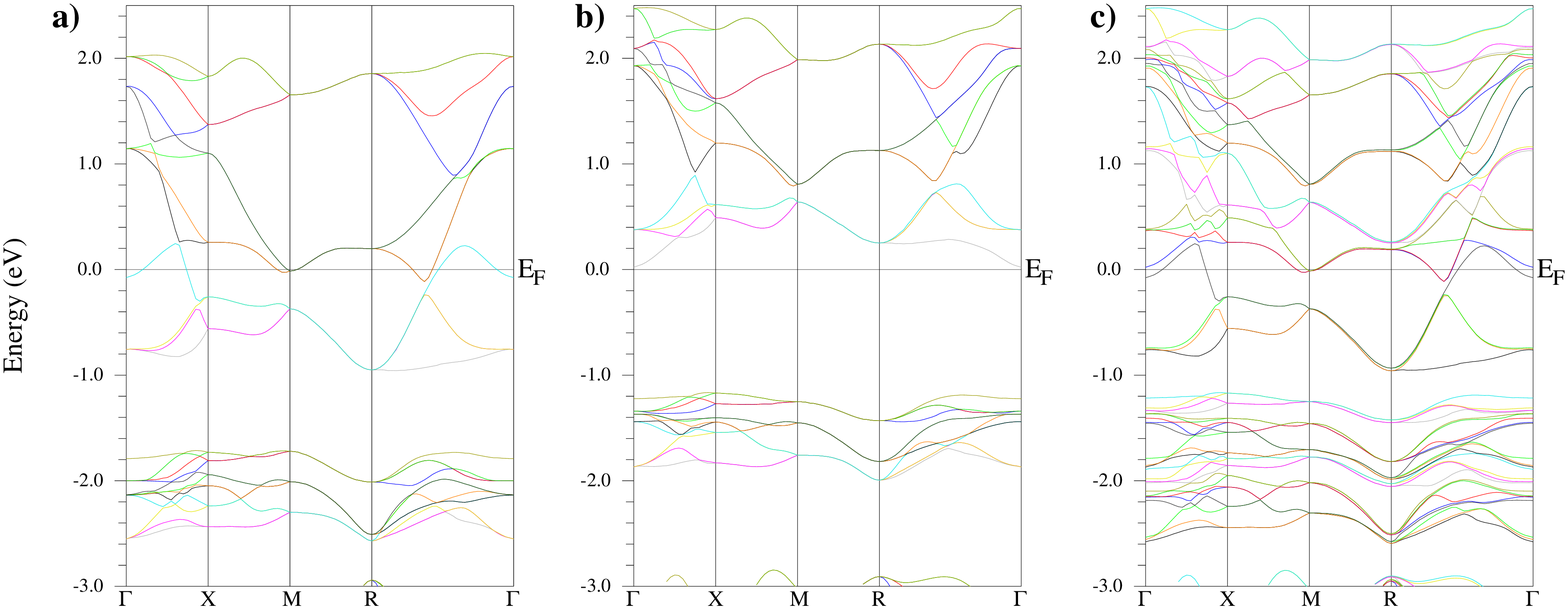}
\caption{(Color online) Band structure plot of the: a) majority spin states, b) minority spin states and c) calculation including spin orbit coupling with the magnetization along 001 direction. Observe that the curvature of the bands remains unaltered when spin-orbit coupling is turned on. Fermi level is at zero}\label{bs}
\end{figure*}

Our results show that a calculation at the GGA level is enough to capture the features of the electronic structure of this itinerant ferromagnet, as shown in a previous work.\cite{martaprb} The introduction of strong correlation effects by the LSDA+U method\cite{sic} does not provide more information about the electronic structure of this itinerant system, in agreement with the work of Kwon \textit{et al}.\cite{kwon} The correct ferromagnetic ground state is obtained at the GGA level (a FM solution is more stable by 75 meV/Co than the paramagnetic phase). A peak of the DOS at the Fermi energy (not shown) is found in a non-magnetic calculation, suggestive of the system being close to a ferromagnetic instability according to the Stoner criterion\cite{stoner} ($N(E_{F})I>1$ to favor ferromagnetism).

Band structure plots (Fig. \ref{bs}a and b for the majority and minority spin channel, respectively) can be used to understand the essential features of the electronic structure of CoS$_2$. Being a half-metal, all the spectral weight around the Fermi level comes from the majority spin channel (Fig. \ref{bs}a). The minority spin channel (Fig. \ref{bs}b) shows a gap of about 1 eV at the Fermi level. A big electron pocket, centered at the $\Gamma$ point, and a large hole pocket appear in the $\Gamma$-R direction. Additional electron pockets  appear near the M point. These results are similar to other band structure plots published using the GGA scheme,\cite{tatsuya} except for the electron pocket centered at the $\Gamma$ point, which is absent in their calculations.

We have also performed calculations including SOC, with the magnetization along the main directions of the cubic structure: (001), (110) and (111). Calculations yield an orbital magnetic moment $0.037 \mu_B$ (quenched by the crystal field\cite{fazekas}), similar to that reported by Antonov \textit{et al}.\cite{antonov} Together with the local spin magnetic moment of $0.92 \mu_B$ (inside the Co muffin tin sphere) we obtain $M_{l}/M_{s}=0.04$, i.e., $<l_{z}>/<s_{z}>=0.08$, in good agreement with experimental measurements ($0.1$).\cite{muro}

A very small energy difference is obtained between the calculations with the magnetization along the three main directions of the crystal, suggesting a very small magnetocrystalline anisotropy in this largely isotropic material. The energy difference between the easy and hard axis is smaller than 50 $\mu$eV/Co atom.

Analyzing the band structure (Fig. \ref{bs}c) obtained from a GGA+SOC (001) calculation, we  observe that the introduction of SOC leads to small shifts in the bands, but the positions of the pockets in the Brillouin zone and the curvature of the bands remain roughly unchanged (we see all the electron and hole pockets discussed above). A splitting in the bands corresponding to the electron pocket in the $\Gamma$-R direction, and a variation in the size of the other pockets can be seen. Below we discuss further the quantitative effects in the electronic structure of introducing SOC.

\begin{figure}
\includegraphics[width=\columnwidth,draft=false]{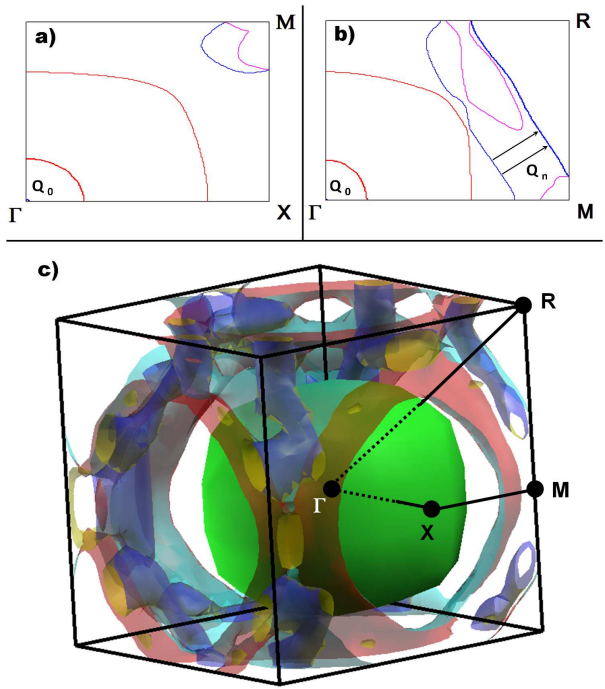}
\caption{(Color online) Fermi surface of the CoS$_2$ obtained for the majority spin channel in a FM solution, within a GGA+SOC-based calculation along the (001) direction. a) and b) show the 2-dimensional Fermi surface in the $\Gamma$-X-M and $\Gamma$-M-R planes, respectively. c) shows the 3-dimensional Fermi surface. In the inset b) we have denoted the nesting vector as Q$_n$ (which corresponds to a wavelength of approximately 6 times the lattice parameter). This would lead to a peak in the spin susceptibility associated with that wave-vector, together with the ferromagnetic ones.}\label{fs}
\end{figure}

For a better description of the states near the Fermi level, Fig. \ref{fs} shows the Fermi surface of the compound, within a GGA+SOC-based calculation with the magnetization set along the (001) direction. In the plot we observe a large isotropic electron pocket around the $\Gamma$ point. Experimentally, \cite{martaprb} the spin susceptibility of this compound can be understood in terms of spin fluctuation theory.\cite{moriya} For this type of systems, peaks in the spin susceptibility can be associated to nesting features in the Fermi surface. In the case of CoS$_2$, some nesting features can be observed in the $\Gamma-M-R$ plane (Fig. \ref{fs} b)), with a nesting vector that we have denoted as $Q_n$. The magnitude of $Q_n$ would correspond to long-wavelength spin fluctuations ($\lambda=6a$, with $a$ being the lattice parameter), that are believed to exist in weak ferromagnets such as CoS$_2$\cite{martaprb} and also MnSi,\cite{fran_mnsi} together with many other systems.\cite{moriya} The rest of the Fermi surface, i.e. the hole pocket along $\Gamma$-R direction and the small electron pockets in the X-M direction near M, is largely incoherent and will lead to negligible nesting.

\subsection{Half-metallicity}

\begin{figure}
\includegraphics[width=\columnwidth,draft=false]{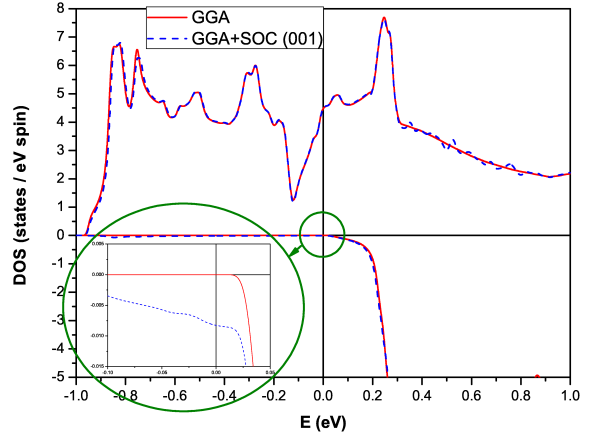}
\caption{(Color online) Spin-up (positive values) and spin-down (negative values) total DOS plots for the CoS$_2$ in the ferromagnetic configuration with a GGA (solid red line) and a GGA+SOC (SOC along (001) direction) (dash blue line). Fermi energy is at zero. The system is half-metallic in the GGA 0 and 1 metallic in the GGA+SOC 0. The inset shows a zoom of the minority DOS near the 0 level of both solutions (GGA and GGA+SOC)}\label{dos_gga_gga+so}
\end{figure}

CoS$_2$ has been predicted to be a good source of spin polarized electrons;\cite{wang} hence, we will try to address this point with our calculations. GGA alone leads to half-metallic behavior (Fig. \ref{dos}), as we saw above. Due to the proximity of  the conduction band edge to the Fermi level in the minority spin channel, SOC destroys the half-metallicity noticeably, as we can see in Fig. \ref{dos_gga_gga+so}. No minority DOS is seen in the GGA calculation, but magnifying the minority DOS of the calculation including SOC (in the inset), we see the DOS is nonzero at the Fermi level for the minority spin channel. This has important consequences for the conduction properties of the material. We can compare the values of the DOS at the Fermi level for the two spin channels, being no longer infinite but N(E$_F)_{\uparrow}$/N(E$_F)_{\downarrow} = 540$ when SOC is considered, giving a polarization of the DOS at the Fermi level $P=[N_{\uparrow}(E_{F})-N_{\downarrow}(E_{F})]/[N_{\uparrow}(E_{F})+N_{\downarrow}(E_{F})] $ of 99.6\%, very large still.

\begin{figure}
\includegraphics[width=\columnwidth,draft=false]{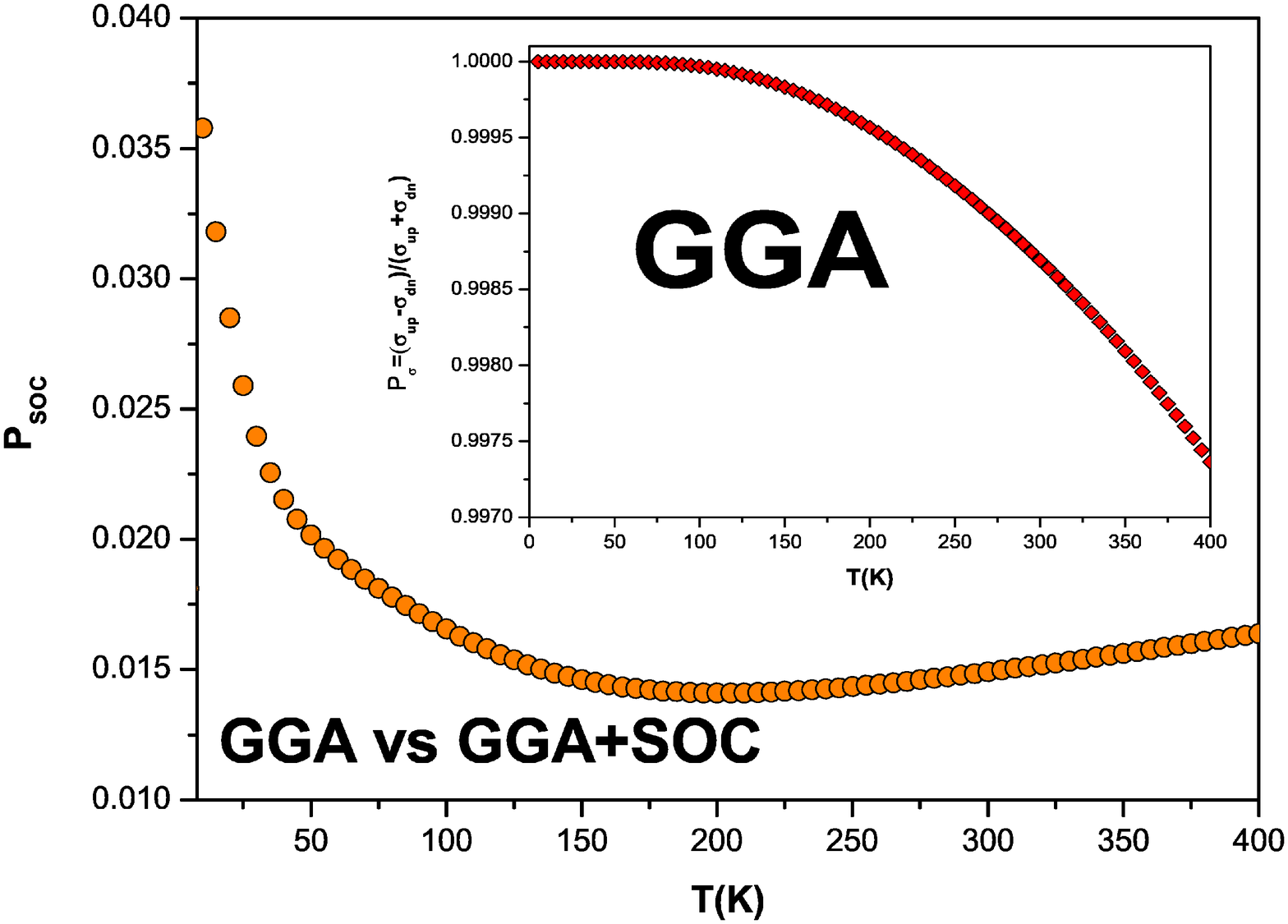}
\caption{(Color online) Percentage (P$_{SOC}$)of the better conduction of GGA versus GGA+SOC calculations. This is calculated by subtraction of the conductivities coming from GGA and GGA+SOC calculations, normalized to the sum of both them (solid orange circles) (P$_{SOC}$= ($\sigma_{no soc}-\sigma_{soc})/(\sigma_{no soc}+\sigma_{soc}$)). GGA alone leads better conduction properties, so the GGA+SOC implies larger resistivity. The inset show the 2 of the conductivities coming from both spin channels over the sum of these conductivities within the GGA 0 (solid red diamonds). At room temperature, $\sigma_{up}$ contributes in a 99.8 \% to the total conductivity and $\sigma_{dn}$ 0.2 \% contributes it to the total conductivity in the GGA 0.}\label{sigma}
\end{figure}

For analyzing further the degree of spin polarization of the carriers in this system, we have calculated the transport properties from our band structure calculations using a semiclassical 1 based on Boltzmann transport theory.\cite{boltztrap} In order to understand the spin polarization of the carriers, it is necessary to go beyond the values of the DOS at the Fermi level. One possible way is to compare the conductivities of both spin channels. The inset of Fig. \ref{sigma} shows the temperature dependence of the ratio between the subtraction and the sum of the conductivities coming from both spin channels, i.e. P$_{\sigma}$= ($\sigma_{up}-\sigma_{dn})/(\sigma_{up}+\sigma_{dn}$), defined analogously to that of the DOS. This quantity ranges from $\pm$1 for a full-polarization to 0 for no polarization and will give us the spin polarization of the carriers at a given temperature. When no SOC is considered and polarization of the DOS is infinite, it is still very large in terms of the conductivities (spin polarization of the carriers), being approximately P$_\sigma$ = 0.998 at room temperature.
 
One would naively expect that the introduction of SOC leads to a substantially larger conductivity coming from the minority spin channel compared to the case without SOC, because of the non-zero DOS at the Fermi level. To gain some insight on that, we present in Fig. \ref{sigma} the temperature dependence of the calculated ratio between the subtraction and the addition of the conductivities calculated with and without SOC: P$_{SOC}$= ($\sigma_{no soc}-\sigma_{soc})/(\sigma_{no soc}+\sigma_{soc}$). We observe that the solution without SOC has a larger conductivity than the solution including SOC up to room temperature, both values differing by about a few percent. Thus, the bigger change in the conductivity when introducing SOC in the calculations is not the additional contribution from the minority spin states, but an overall reduction in the conductivity. Even though a separation in conductivity from both spin channels is not possible when calculating with SOC, the results suggest a very large polarization of the carriers will still be present when the calculations include SOC. This effect will account for a reduction in the half-metallicity of just a few percent. However, to understand the experimental observations, one needs to consider other effects as possible sources against a full spin polarization, such as defects, spin excitations or non-quasiparticle states\cite{ebert, chioncel} to be important.

%. This means larger resititivity introducing spin orbit coupling. Because of the shape of the bands crossing the Fermi level, and the 1 of $\sigma$ on that, it may be the principal reason to obtain the negative MR in the compound. 0 effects of \textit{LS} 0 make non zero 0 on the minority spin channel and increase the conductivitity coming from $\sigma_{\downarrow}$, and temperature and applied magnetic field increase the population on the majority spin channel leading large difference on the conductivities between two spin channels ($\sigma_{\uparrow}$ and $\sigma_{\downarrow}$). 

\begin{figure}
\includegraphics[width=\columnwidth,draft=false]{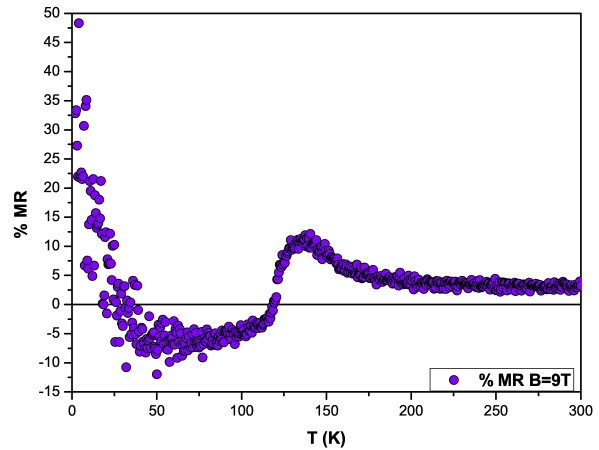}
\caption{(Color online) Evolution of the magnetoresistance with temperature measured comparing the values of resistivity at $H=9T$ and $H=0T$ applied magnetic field according to the formula MR=$100(\rho_{9T}-\rho_{0T})/\rho_{0T}$. Below 25K we observed positive MR, and negative MR are obtained in the rest of FM phase.}\label{experimental_MR}
\end{figure}

In a material with such a large polarization of carriers at the Fermi level, a negative magnetoresistance is expected to occur in the FM region of the phase diagram. Figure \ref{experimental_MR} shows the measurements of the magnetoresistace (MR) in an external magnetic field of 9 T, (MR=$100(\rho_{9T}-\rho_{0T})/\rho_{0T}$). These show a negative MR in a temperature range within the FM phase (below the Curie temperature $T_c=122K$\cite{martaprb}) with a value ranging between -3\% and -9\%. At low temperatures, a large positive value of the MR is observed and, above the Curie point, in the paramagnetic phase, the MR becomes  again positive, approaching a constant value of about 3$\%$ at high temperatures. MR does not reach a very high value, consistent with the fact that spin polarization of the carriers is not complete, as discussed above.

%Obtaining both spin channel contributions separately for the conductivity including SOC is not possible. Hence, we use the bandstructure without SOC and approximate the effects of SOC as a rigid shift to the bands, such that half-metallicity is broken. As we saw above, this is a very good approximation to the effects of SOC in CoS$_2$. For doing this, we just move the Fermi energy for both spin channels so that the calculation without including SOC matches the DOS at the Fermi energy calculated with SOC, and then we calculate the conductivities from those displaced bands that try to mimic the effects of SOC in the band structure.

%We observe that introducing SOC, even though it destroys the half-metallicity, does not destroy substantially the spin polarization of the carriers. Previous ab initio works on the issue do not consider the effects of SOC in the conductivity. In this work, even though we show they are 0 in terms of the electronic structure, a large spin polarization of the carriers is also expected.

\section{Summary}

We have analyzed the electronic structure and magnetic properties of this itinerant weak ferromagnet, dominated by spin-density fluctuations. We have related these to the nesting features in the Fermi surface associated to long wavelength nesting vectors. Being an itinerant system, the GGA scheme can describe its properties and electronic structure accurately, not needing to introduce the electron correlations with means of LDA+U.
GGA calculations alone predict CoS$_{2}$ to be half metallic, but when introducing the important effects of SOC effects, the results of the calculations show SOC destroys the half-metallicity, non-zero occupation (band-crossing) of the minority channel at the Fermi level appear. Our transport properties calculations show that changes in the conductivity associated to the introduction of SOC in the calculations do not imply a significant break-up of the half-metallic behavior, but just a few percent. These results are consistent with our experimental observation of negative magnetoresistance in the material.

\section{Acknowledgments}

F. Rivadulla is acknowledged for growing and characterizing the samples of CoS$_2$ and measuring magnetoresistance shown in Fig. \ref{experimental_MR}. J. Castro is acknowledged for fruitful discussions. M. Pereiro and J. Botana are acknowledged for their technical support. The authors thank the CESGA (Centro de Supercomputacion de Galicia) for the computing facilities and the Ministerio de Educaci\'on y Ciencia (MEC)  for the financial support through the project MAT2009-08165. A.S.B. thanks MEC for an FPU grant. We are also thankful to the Xunta de Galicia for financial support through the project INCITE08PXIB236053PR.


\begin{thebibliography}{36}

\expandafter\ifx\csname natexlab\endcsname\relax\def\natexlab#1{#1}\fi
\expandafter\ifx\csname bibnamefont\endcsname\relax
  \def\bibnamefont#1{#1}\fi
\expandafter\ifx\csname bibfnamefont\endcsname\relax
  \def\bibfnamefont#1{#1}\fi
\expandafter\ifx\csname citenamefont\endcsname\relax
  \def\citenamefont#1{#1}\fi
\expandafter\ifx\csname url\endcsname\relax
  \def\url#1{\texttt{#1}}\fi
\expandafter\ifx\csname urlprefix\endcsname\relax\def\urlprefix{URL }\fi
\providecommand{\bibinfo}[2]{#2}
\providecommand{\eprint}[2][]{\url{#2}}

\bibitem[{\citenamefont{Katsnelson et~al.}(2008)\citenamefont{Katsnelson,
  Irkhin, Chioncel, Lichtenstein, and de~Groot}}]{katsnelson}
\bibinfo{author}{\bibfnamefont{M.~I.} \bibnamefont{Katsnelson}},
  \bibinfo{author}{\bibfnamefont{V.~Y.} \bibnamefont{Irkhin}},
  \bibinfo{author}{\bibfnamefont{L.}~\bibnamefont{Chioncel}},
  \bibinfo{author}{\bibfnamefont{A.~I.} \bibnamefont{Lichtenstein}},
  \bibnamefont{and} \bibinfo{author}{\bibfnamefont{R.~A.}
  \bibnamefont{de~Groot}}, \bibinfo{journal}{Rev. Mod. Phys.}
  \textbf{\bibinfo{volume}{80}}, \bibinfo{pages}{315} (\bibinfo{year}{2008}).

\bibitem[{\citenamefont{Schwarz}(1986)}]{schwarz_half}
\bibinfo{author}{\bibfnamefont{K.}~\bibnamefont{Schwarz}}, \bibinfo{journal}{J.
  Phys. F: Met. Phys.} \textbf{\bibinfo{volume}{16}}, \bibinfo{pages}{L211}
  (\bibinfo{year}{1986}).

\bibitem[{\citenamefont{Wolf et~al.}(2001)\citenamefont{Wolf, Awschalom,
  Buhrman, Daughton, von Moln\'{a}r, Roukes, Chtchelkanova, and
  Treger}}]{spintronic}
\bibinfo{author}{\bibfnamefont{S.}~\bibnamefont{Wolf}},
  \bibinfo{author}{\bibfnamefont{D.}~\bibnamefont{Awschalom}},
  \bibinfo{author}{\bibfnamefont{R.}~\bibnamefont{Buhrman}},
  \bibinfo{author}{\bibfnamefont{J.}~\bibnamefont{Daughton}},
  \bibinfo{author}{\bibfnamefont{S.}~\bibnamefont{von Moln\'{a}r}},
  \bibinfo{author}{\bibfnamefont{M.}~\bibnamefont{Roukes}},
  \bibinfo{author}{\bibfnamefont{A.}~\bibnamefont{Chtchelkanova}},
  \bibnamefont{and} \bibinfo{author}{\bibfnamefont{D.}~\bibnamefont{Treger}},
  \bibinfo{journal}{Science} \textbf{\bibinfo{volume}{294}},
  \bibinfo{pages}{1488} (\bibinfo{year}{2001}).

\bibitem[{\citenamefont{Zhao et~al.}(1993)\citenamefont{Zhao, Callaway, and
  Hayashibara}}]{zhao}
\bibinfo{author}{\bibfnamefont{G.~L.} \bibnamefont{Zhao}},
  \bibinfo{author}{\bibfnamefont{J.}~\bibnamefont{Callaway}}, \bibnamefont{and}
  \bibinfo{author}{\bibfnamefont{M.}~\bibnamefont{Hayashibara}},
  \bibinfo{journal}{Phys. Rev. B} \textbf{\bibinfo{volume}{48}},
  \bibinfo{pages}{15781} (\bibinfo{year}{1993}).

\bibitem[{\citenamefont{Mazin}(2000)}]{mazin}
\bibinfo{author}{\bibfnamefont{I.~I.} \bibnamefont{Mazin}},
  \bibinfo{journal}{Appl. Phys. Lett.} \textbf{\bibinfo{volume}{77}},
  \bibinfo{pages}{3000} (\bibinfo{year}{2000}).

\bibitem[{\citenamefont{Kwon et~al.}(2000)\citenamefont{Kwon, Youn, and
  Min}}]{kwon}
\bibinfo{author}{\bibfnamefont{S.~K.} \bibnamefont{Kwon}},
  \bibinfo{author}{\bibfnamefont{S.~J.} \bibnamefont{Youn}}, \bibnamefont{and}
  \bibinfo{author}{\bibfnamefont{B.~I.} \bibnamefont{Min}},
  \bibinfo{journal}{Phys. Rev. B} \textbf{\bibinfo{volume}{62}},
  \bibinfo{pages}{357} (\bibinfo{year}{2000}).

\bibitem[{\citenamefont{Shishidou et~al.}(2001)\citenamefont{Shishidou,
  Freeman, and Asahi}}]{tatsuya}
\bibinfo{author}{\bibfnamefont{T.}~\bibnamefont{Shishidou}},
  \bibinfo{author}{\bibfnamefont{A.~J.} \bibnamefont{Freeman}},
  \bibnamefont{and} \bibinfo{author}{\bibfnamefont{R.}~\bibnamefont{Asahi}},
  \bibinfo{journal}{Phys. Rev. B} \textbf{\bibinfo{volume}{64}},
  \bibinfo{pages}{180401(R)} (\bibinfo{year}{2001}).

\bibitem[{\citenamefont{Jin and Lee}(2006)}]{jiu_jin}
\bibinfo{author}{\bibfnamefont{Y.~J.} \bibnamefont{Jin}} \bibnamefont{and}
  \bibinfo{author}{\bibfnamefont{J.~I.} \bibnamefont{Lee}},
  \bibinfo{journal}{Phys. Rev. B} \textbf{\bibinfo{volume}{73}},
  \bibinfo{pages}{064405} (\bibinfo{year}{2006}).

\bibitem[{\citenamefont{Wu et~al.}(2007)\citenamefont{Wu, Losovyj, Wisbey,
  Belashchenko, Manno, Wang, Leighton, and Dowben}}]{ning_wu}
\bibinfo{author}{\bibfnamefont{N.}~\bibnamefont{Wu}},
  \bibinfo{author}{\bibfnamefont{Y.~B.} \bibnamefont{Losovyj}},
  \bibinfo{author}{\bibfnamefont{D.}~\bibnamefont{Wisbey}},
  \bibinfo{author}{\bibfnamefont{K.}~\bibnamefont{Belashchenko}},
  \bibinfo{author}{\bibfnamefont{M.}~\bibnamefont{Manno}},
  \bibinfo{author}{\bibfnamefont{L.}~\bibnamefont{Wang}},
  \bibinfo{author}{\bibfnamefont{C.}~\bibnamefont{Leighton}}, \bibnamefont{and}
  \bibinfo{author}{\bibfnamefont{P.~A.} \bibnamefont{Dowben}},
  \bibinfo{journal}{J. Phys.: Condens. Matter} \textbf{\bibinfo{volume}{19}},
  \bibinfo{pages}{156224} (\bibinfo{year}{2007}).

\bibitem[{\citenamefont{de~Jong and Beenakker}(1995)}]{de_jong}
\bibinfo{author}{\bibfnamefont{M.~J.~M.} \bibnamefont{de~Jong}}
  \bibnamefont{and} \bibinfo{author}{\bibfnamefont{C.~W.~J.}
  \bibnamefont{Beenakker}}, \bibinfo{journal}{Phys. Rev. Lett.}
  \textbf{\bibinfo{volume}{74}}, \bibinfo{pages}{1657} (\bibinfo{year}{1995}).

\bibitem[{\citenamefont{Wang et~al.}(2004)\citenamefont{Wang, Chen, and
  Leighton}}]{wang}
\bibinfo{author}{\bibfnamefont{L.}~\bibnamefont{Wang}},
  \bibinfo{author}{\bibfnamefont{T.~Y.} \bibnamefont{Chen}}, \bibnamefont{and}
  \bibinfo{author}{\bibfnamefont{C.}~\bibnamefont{Leighton}},
  \bibinfo{journal}{Phys. Rev. B} \textbf{\bibinfo{volume}{69}},
  \bibinfo{pages}{094412} (\bibinfo{year}{2004}).

\bibitem[{\citenamefont{Wang et~al.}(2005)\citenamefont{Wang, Umemoto,
  Wentzcovitch, Chen, Chien, Checkelsky, Eckert, Dahlberg, and
  Leighton}}]{wang_2}
\bibinfo{author}{\bibfnamefont{L.}~\bibnamefont{Wang}},
  \bibinfo{author}{\bibfnamefont{K.}~\bibnamefont{Umemoto}},
  \bibinfo{author}{\bibfnamefont{R.~M.} \bibnamefont{Wentzcovitch}},
  \bibinfo{author}{\bibfnamefont{T.~Y.} \bibnamefont{Chen}},
  \bibinfo{author}{\bibfnamefont{C.~L.} \bibnamefont{Chien}},
  \bibinfo{author}{\bibfnamefont{J.~G.} \bibnamefont{Checkelsky}},
  \bibinfo{author}{\bibfnamefont{J.~C.} \bibnamefont{Eckert}},
  \bibinfo{author}{\bibfnamefont{E.~D.} \bibnamefont{Dahlberg}},
  \bibnamefont{and} \bibinfo{author}{\bibfnamefont{C.}~\bibnamefont{Leighton}},
  \bibinfo{journal}{Phys. Rev. Lett.} \textbf{\bibinfo{volume}{94}},
  \bibinfo{pages}{056602} (\bibinfo{year}{2005}).

\bibitem[{\citenamefont{Brown et~al.}(2005)\citenamefont{Brown, Neumann, Simon,
  Ueno, and Ziebeck}}]{brown}
\bibinfo{author}{\bibfnamefont{P.~J.} \bibnamefont{Brown}},
  \bibinfo{author}{\bibfnamefont{K.-U.} \bibnamefont{Neumann}},
  \bibinfo{author}{\bibfnamefont{A.}~\bibnamefont{Simon}},
  \bibinfo{author}{\bibfnamefont{F.}~\bibnamefont{Ueno}}, \bibnamefont{and}
  \bibinfo{author}{\bibfnamefont{K.~R.~A.} \bibnamefont{Ziebeck}},
  \bibinfo{journal}{J. Phys.: Condens. Matter} \textbf{\bibinfo{volume}{17}},
  \bibinfo{pages}{1583} (\bibinfo{year}{2005}).

\bibitem[{\citenamefont{Takahashi et~al.}(2001)\citenamefont{Takahashi, Naitoh,
  Sato, Kamiyama, Yamada, Hiraka, Endoh, Usuda, and Hamada}}]{takahashi}
\bibinfo{author}{\bibfnamefont{T.}~\bibnamefont{Takahashi}},
  \bibinfo{author}{\bibfnamefont{Y.}~\bibnamefont{Naitoh}},
  \bibinfo{author}{\bibfnamefont{T.}~\bibnamefont{Sato}},
  \bibinfo{author}{\bibfnamefont{T.}~\bibnamefont{Kamiyama}},
  \bibinfo{author}{\bibfnamefont{K.}~\bibnamefont{Yamada}},
  \bibinfo{author}{\bibfnamefont{H.}~\bibnamefont{Hiraka}},
  \bibinfo{author}{\bibfnamefont{Y.}~\bibnamefont{Endoh}},
  \bibinfo{author}{\bibfnamefont{M.}~\bibnamefont{Usuda}}, \bibnamefont{and}
  \bibinfo{author}{\bibfnamefont{N.}~\bibnamefont{Hamada}},
  \bibinfo{journal}{Phys. Rev. B} \textbf{\bibinfo{volume}{63}},
  \bibinfo{pages}{094415} (\bibinfo{year}{2001}).

\bibitem[{\citenamefont{Otero-Leal et~al.}(2008)\citenamefont{Otero-Leal,
  Rivadulla, Garc\'{i}a-Hern\'{a}ndez, neiro, Pardo, Baldomir, and
  Rivas}}]{martaprb}
\bibinfo{author}{\bibfnamefont{M.}~\bibnamefont{Otero-Leal}},
  \bibinfo{author}{\bibfnamefont{F.}~\bibnamefont{Rivadulla}},
  \bibinfo{author}{\bibfnamefont{M.}~\bibnamefont{Garc\'{i}a-Hern\'{a}ndez}},
  \bibinfo{author}{\bibfnamefont{A.}~\bibnamefont{Pi\~neiro}},
  \bibinfo{author}{\bibfnamefont{V.}~\bibnamefont{Pardo}},
  \bibinfo{author}{\bibfnamefont{D.}~\bibnamefont{Baldomir}}, \bibnamefont{and}
  \bibinfo{author}{\bibfnamefont{J.}~\bibnamefont{Rivas}},
  \bibinfo{journal}{Phys. Rev. B} \textbf{\bibinfo{volume}{78}},
  \bibinfo{pages}{180415(R)} (\bibinfo{year}{2008}).

\bibitem[{\citenamefont{Perdew et~al.}(1996)\citenamefont{Perdew, Burke, and
  Ernzerhof}}]{gga}
\bibinfo{author}{\bibfnamefont{J.~P.}~\bibnamefont{Perdew}},
  \bibinfo{author}{\bibfnamefont{K.}~\bibnamefont{Burke}}, \bibnamefont{and}
  \bibinfo{author}{\bibfnamefont{M.}~\bibnamefont{Ernzerhof}},
  \bibinfo{journal}{Phys.\ Rev. Lett.} \textbf{\bibinfo{volume}{77}},
  \bibinfo{pages}{3865} (\bibinfo{year}{1996}).

\bibitem[{\citenamefont{Mavropoulos et~al.}(2004)\citenamefont{Mavropoulos,
  Sato, Zeller, Dederichs, Popescu, and Ebert}}]{mavropoulos}
\bibinfo{author}{\bibfnamefont{P.}~\bibnamefont{Mavropoulos}},
  \bibinfo{author}{\bibfnamefont{K.}~\bibnamefont{Sato}},
  \bibinfo{author}{\bibfnamefont{R.}~\bibnamefont{Zeller}},
  \bibinfo{author}{\bibfnamefont{P.~H.}~\bibnamefont{Dederichs}},
  \bibinfo{author}{\bibfnamefont{V.}~\bibnamefont{Popescu}}, \bibnamefont{and}
  \bibinfo{author}{\bibfnamefont{H.}~\bibnamefont{Ebert}},
  \bibinfo{journal}{Phys. Rev. B} \textbf{\bibinfo{volume}{69}},
  \bibinfo{pages}{054424} (\bibinfo{year}{2004}).

\bibitem[{\citenamefont{Pardo and Pickett}(2009)}]{pardo_hmafm}
\bibinfo{author}{\bibfnamefont{V.}~\bibnamefont{Pardo}} \bibnamefont{and}
  \bibinfo{author}{\bibfnamefont{W.~E.} \bibnamefont{Pickett}},
  \bibinfo{journal}{Phys. Rev. B} \textbf{\bibinfo{volume}{80}},
  \bibinfo{pages}{054415} (\bibinfo{year}{2009}).

\bibitem[{\citenamefont{Hohenberg and Kohn}(1964)}]{dft}
\bibinfo{author}{\bibfnamefont{P.}~\bibnamefont{Hohenberg}} \bibnamefont{and}
  \bibinfo{author}{\bibfnamefont{W.}~\bibnamefont{Kohn}},
  \bibinfo{journal}{Phys. Rev.} \textbf{\bibinfo{volume}{136}},
  \bibinfo{pages}{B864} (\bibinfo{year}{1964}).

\bibitem[{\citenamefont{Schwarz and Blaha}(2003)}]{wien}
\bibinfo{author}{\bibfnamefont{K.}~\bibnamefont{Schwarz}} \bibnamefont{and}
  \bibinfo{author}{\bibfnamefont{P.}~\bibnamefont{Blaha}},
  \bibinfo{journal}{Comp. Mat. Sci.} \textbf{\bibinfo{volume}{28}},
  \bibinfo{pages}{259} (\bibinfo{year}{2003}).

\bibitem[{\citenamefont{Blaha et~al.}(2001)\citenamefont{Blaha, Schwarz,
  Madsen, Kvasnicka, and Luitz}}]{wien2k}
\bibinfo{author}{\bibfnamefont{P.}~\bibnamefont{Blaha}},
  \bibinfo{author}{\bibfnamefont{K.}~\bibnamefont{Schwarz}},
  \bibinfo{author}{\bibfnamefont{G.~K.~H.} \bibnamefont{Madsen}},
  \bibinfo{author}{\bibfnamefont{D.}~\bibnamefont{Kvasnicka}},
  \bibnamefont{and} \bibinfo{author}{\bibfnamefont{J.}~\bibnamefont{Luitz}},
  \emph{\bibinfo{title}{WIEN2k, An Augmented Plane Wave Plus Local Orbitals
  Program for Calculating Crystal Properties. ISBN 3-9501031-1-2}},
  \bibinfo{address}{Vienna University of Technology, Austria}
  (\bibinfo{year}{2001}).

\bibitem[{\citenamefont{Sj{\"o}stedt et~al.}(2000)\citenamefont{Sj{\"o}stedt,
  N{\"o}rdstrom, and Singh}}]{sjo}
\bibinfo{author}{\bibfnamefont{E.}~\bibnamefont{Sj{\"o}stedt}},
  \bibinfo{author}{\bibfnamefont{L.}~\bibnamefont{N{\"o}rdstrom}},
  \bibnamefont{and} \bibinfo{author}{\bibfnamefont{D.}~\bibnamefont{Singh}},
  \bibinfo{journal}{Solid State Commun.} \textbf{\bibinfo{volume}{114}},
  \bibinfo{pages}{15} (\bibinfo{year}{2000}).

\bibitem[{\citenamefont{Madsen and Singh}(2006)}]{boltztrap}
\bibinfo{author}{\bibfnamefont{G.~K.~H.} \bibnamefont{Madsen}}
  \bibnamefont{and} \bibinfo{author}{\bibfnamefont{D.~J.} \bibnamefont{Singh}},
  \bibinfo{journal}{Comp. Phys. Commun.} \textbf{\bibinfo{volume}{175}},
  \bibinfo{pages}{67} (\bibinfo{year}{2006}).

\bibitem[{\citenamefont{Singh}(1994)}]{singh}
\bibinfo{author}{\bibfnamefont{D.}~\bibnamefont{Singh}},
  \emph{\bibinfo{title}{Planewaves, pseudopotentials and LAPW method}}
  (\bibinfo{address}{Kluwer Academic Publishers}, \bibinfo{year}{1994}).

\bibitem[{\citenamefont{Wickoff}(1965)}]{wyck}
\bibinfo{author}{\bibfnamefont{R.~W.~G.} \bibnamefont{Wickoff}},
  \emph{\bibinfo{title}{Crystal Structures, Vol. 1}}
  (\bibinfo{address}{Interscience, New York}, \bibinfo{year}{1965}).

\bibitem[{\citenamefont{Yamamoto et~al.}(1999)\citenamefont{Yamamoto, Machida,
  Moritomo, and Nakamura}}]{yamamoto}
\bibinfo{author}{\bibfnamefont{R.}~\bibnamefont{Yamamoto}},
  \bibinfo{author}{\bibfnamefont{A.}~\bibnamefont{Machida}},
  \bibinfo{author}{\bibfnamefont{Y.}~\bibnamefont{Moritomo}}, \bibnamefont{and}
  \bibinfo{author}{\bibfnamefont{A.}~\bibnamefont{Nakamura}},
  \bibinfo{journal}{Phys. Rev. B} \textbf{\bibinfo{volume}{59}},
  \bibinfo{pages}{R7793} (\bibinfo{year}{1999}).

\bibitem[{\citenamefont{Kokalj}(1999)}]{xcrysden}
\bibinfo{author}{\bibfnamefont{A.}~\bibnamefont{Kokalj}}, \bibinfo{journal}{J.
  Mol. Graphics Modell.} \textbf{\bibinfo{volume}{17}}, \bibinfo{pages}{176}
  (\bibinfo{year}{1999}).

\bibitem[{\citenamefont{Lichtenstein et~al.}(1995)\citenamefont{Lichtenstein,
  Anisimov, and Zaanen}}]{sic}
\bibinfo{author}{\bibfnamefont{A.~I.}~\bibnamefont{Lichtenstein}},
  \bibinfo{author}{\bibfnamefont{V.~I.}~\bibnamefont{Anisimov}}, \bibnamefont{and}
  \bibinfo{author}{\bibfnamefont{J.}~\bibnamefont{Zaanen}},
  \bibinfo{journal}{Phys.\ Rev. B} \textbf{\bibinfo{volume}{52}},
  \bibinfo{pages}{R5467} (\bibinfo{year}{1995}).

\bibitem[{\citenamefont{Stoner}(1938)}]{stoner}
\bibinfo{author}{\bibfnamefont{E.~S.} \bibnamefont{Stoner}},
  \bibinfo{journal}{Proc. R. Soc.} \textbf{\bibinfo{volume}{165}},
  \bibinfo{pages}{372} (\bibinfo{year}{1938}).

\bibitem[{\citenamefont{Fazekas}(1999)}]{fazekas}
\bibinfo{author}{\bibfnamefont{P.}~\bibnamefont{Fazekas}},
  \emph{\bibinfo{title}{Lecture Notes on Electron Correlation and Magnetism}}
  (\bibinfo{address}{World Scientific Publishing Co. Pte. Ltd., Singapore},
  \bibinfo{year}{1999}).

\bibitem[{\citenamefont{Antonov et~al.}(2008)\citenamefont{Antonov,
  Andryushchenko, Shpak, Yaresko, and Jepsen}}]{antonov}
\bibinfo{author}{\bibfnamefont{V.~N.} \bibnamefont{Antonov}},
  \bibinfo{author}{\bibfnamefont{O.~V.} \bibnamefont{Andryushchenko}},
  \bibinfo{author}{\bibfnamefont{A.~P.} \bibnamefont{Shpak}},
  \bibinfo{author}{\bibfnamefont{A.~N.} \bibnamefont{Yaresko}},
  \bibnamefont{and} \bibinfo{author}{\bibfnamefont{O.}~\bibnamefont{Jepsen}},
  \bibinfo{journal}{Phys. Rev. B} \textbf{\bibinfo{volume}{78}},
  \bibinfo{pages}{094409} (\bibinfo{year}{2008}).

\bibitem[{\citenamefont{Muro et~al.}(1996)\citenamefont{Muro, Shishidou, Oda,
  Fukawa, Yamada, Kimura, Imada, Suga, Park, Miyahara and Sato}}]{muro}
\bibinfo{author}{\bibfnamefont{T.}~\bibnamefont{Muro}},
  \bibinfo{author}{\bibfnamefont{T.}~\bibnamefont{Shishidou}},
  \bibinfo{author}{\bibfnamefont{F.}~\bibnamefont{Oda}},
  \bibinfo{author}{\bibfnamefont{T.}~\bibnamefont{Fukawa}},
  \bibinfo{author}{\bibfnamefont{H.}~\bibnamefont{Yamada}},
  \bibinfo{author}{\bibfnamefont{A.}~\bibnamefont{Kimura}},
  \bibinfo{author}{\bibfnamefont{S.}~\bibnamefont{Imada}}, 
  \bibinfo{author}{\bibfnamefont{S.}~\bibnamefont{Suga}},
  \bibinfo{author}{\bibfnamefont{S.~Y.}~\bibnamefont{Park}},
  \bibinfo{author}{\bibfnamefont{T.}~\bibnamefont{Miyahara}}, \bibnamefont{and}
  \bibinfo{author}{\bibfnamefont{K.}~\bibnamefont{Sato}}, 
  \bibinfo{journal}{Phys. Rev. B} \textbf{\bibinfo{volume}{53}},
  \bibinfo{pages}{7055} (\bibinfo{year}{1996}).

\bibitem[{\citenamefont{Moriya}(1985)}]{moriya}
\bibinfo{author}{\bibfnamefont{T.}~\bibnamefont{Moriya}},
  \emph{\bibinfo{title}{Spin Fluctuations in Itinerant Electron Magnetism}}
  (\bibinfo{address}{Springer-Verlag, Berlin}, \bibinfo{year}{1985}).

\bibitem[{\citenamefont{Otero-Leal et~al.}(2009)\citenamefont{Otero-Leal,
  Rivadulla, Saxena, Ahilan, and Rivas}}]{fran_mnsi}
\bibinfo{author}{\bibfnamefont{M.}~\bibnamefont{Otero-Leal}},
  \bibinfo{author}{\bibfnamefont{F.}~\bibnamefont{Rivadulla}},
  \bibinfo{author}{\bibfnamefont{S.~S.} \bibnamefont{Saxena}},
  \bibinfo{author}{\bibfnamefont{K.}~\bibnamefont{Ahilan}}, \bibnamefont{and}
  \bibinfo{author}{\bibfnamefont{J.}~\bibnamefont{Rivas}},
  \bibinfo{journal}{Phys. Rev. B} \textbf{\bibinfo{volume}{79}},
  \bibinfo{pages}{060401(R)} (\bibinfo{year}{2009}).

\bibitem[{\citenamefont{Ebert and Schütz}(1991)}]{ebert}
\bibinfo{author}{\bibfnamefont{H.}~\bibnamefont{Ebert}} \bibnamefont{and}
  \bibinfo{author}{\bibfnamefont{G.}~\bibnamefont{Schütz}},
  \bibinfo{journal}{J. Appl. Phys.} \textbf{\bibinfo{volume}{69}},
  \bibinfo{pages}{4627} (\bibinfo{year}{1991}).

\bibitem[{\citenamefont{Chioncel et~al.}(2003)\citenamefont{Chioncel,
  Katsnelson, de~Groot, and Lichtenstein}}]{chioncel}
\bibinfo{author}{\bibfnamefont{L.}~\bibnamefont{Chioncel}},
  \bibinfo{author}{\bibfnamefont{M.~I.}~\bibnamefont{Katsnelson}},
  \bibinfo{author}{\bibfnamefont{R.~A.}~\bibnamefont{de~Groot}}, \bibnamefont{and}
  \bibinfo{author}{\bibfnamefont{A.~I.}~\bibnamefont{Lichtenstein}},
  \bibinfo{journal}{Phys. Rev. B} \textbf{\bibinfo{volume}{68}},
  \bibinfo{pages}{144425} (\bibinfo{year}{2003}).

\end{thebibliography}
\end{document}